\documentclass[12pt,a4paper]{article}
\usepackage{amsmath}
\usepackage{amsfonts}
\usepackage{amssymb}
\usepackage{amsthm}

\input{epsf}             % encapsulated Postscript figures
\newcommand{\SU}{{\rm SU}}

\newcommand{\U}{{\rm U}}
\newcommand{\dd}{{\mathrm d}}      % \d already defined
\newcommand{\tr}{{\rm tr}}

\newcommand\Ei{\mbox{Ei}}

\newcommand{\Atop}[2]{\genfrac{}{}{0pt}{}{#1}{#2}} %% fraction without the line

\theoremstyle{definition}
 
\begin{document}
\title{State sum models, induced gravity and the spectral action}

\author{John W. Barrett
%\thanks{Copyright \copyright\ John W. Barrett 2008}
\\ \\
School of Mathematical Sciences\\
University of Nottingham\\
University Park\\
Nottingham NG7 2RD, UK\\
\\
E-mail john.barrett@nottingham.ac.uk}

%\date{1 October 2009}

\maketitle
\begin{abstract} A proposal that the bosonic action of gravity and the standard model is induced from the fermionic action is investigated. It is suggested that this might occur naturally in state sum models.
\end{abstract}

\section{Introduction}
This article concerns the problem of constructing a quantum theory of gravity coupled to matter. Many approaches to quantum gravity assume that the main difficulty lies in reconciling the dynamics of general relativity with quantum mechanics, and therefore start by describing a purely gravitational theory without matter fields. This article however takes the
 opposite point of view: that the gravity-matter interaction is in fact the most important feature of the dynamics. The simplest hypothesis is that the dynamics of the gravitational field itself is a side-effect of the gravity-matter interaction.

At energies below the Planck scale it should be possible to describe the theory to a good approximation with continuum fields and an effective action. In the framework due to Alain Connes\cite{C}, the fields of the standard model and gravity are packaged very effectively into geometry and matter on a certain non-commutative Kaluza-Klein space consisting of the product of a standard (commutative) four-manifold and a non-commutative `internal space'.
 
In this article a new proposal is given for the action of gravity coupled to standard model matter at high energies near (but not exceeding) the Planck scale. This is done by generalising Sakharov's idea of induced gravity in which the gravitational terms in the action are induced by matter interactions \cite{SK}.\footnote{A somewhat different proposal where the gauge bosons are collective modes in a fermion system is made in \cite{KV}.} The generalisation is to propose an induced action for all the bosonic fields in Connes' framework. For this idea to work, it is necessary for the space-time geometry to exhibit discreteness at the Planck scale, so that there is a natural cut-off at the corresponding energy. 

Implementing this proposal by defining a functional integral over the matter and gravitational fields with the required discreteness requires some non-trivial mathematical framework.
It is noted that the main qualitative features that are required for the bosonic part of the functional integral are found in the construction of various topological gauge theories using the techniques of higher category theory and state sum models. However, as yet a model which realises all of the requirements remains to be defined, and in particular it is not yet known how to code the standard model geometry into state sum models. Some general perspectives are offered for a programme in which this might be realised.

\section{Induced gravity and the spectral action}

The induced gravity phenomenon occurs within the framework of Lagrangian quantum field theory. In the form discussed here, it is assumed that there is a fundamental cut-off in the energy scale of the fields, and that the Lagrangian picture is an effective theory below that energy scale (this energy scale will be called the `Planck scale', even though it may not be exactly at the Planck energy). 

In fact one should assume that it isn't necessary for a Lagrangian description to exist at any higher energies, the physics being described in some other completely different mathematical framework. Some proposals for a possible such alternative mathematical framework are made in the next section, but the considerations here would be also valid in other frameworks.

A model problem in which the induced gravity phenomenon occurs is as follows. Consider the fields on the space-time manifold to be a $-+++$ signature metric $g$ and $N$ fermion fields. Let $D_G$ be the gravitational Dirac operator. Then the induced gravity proposal for the action of the quantum field theory is
$$I=\int\left(\overline\psi D_G\psi - 2\Lambda_0\right)\dd V.$$
The functional integral which gives the quantum amplitudes of the theory is then, heuristically,
$$Z=\int e^{iI}\; \dd g\,\dd\psi\,\dd\bar\psi.$$
Note there is no Einstein-Hilbert term although the metric is integrated over all fluctuations of the metric. The fermion integration is a Gaussian, so can be defined precisely once a high-energy cutoff is specified. However the definition of the integration over all metrics requires further mathematical technology, for example, the state sums considered further below.

%Let the cut-off energy be $c$, and $D_c$ the Dirac operator with this cutoff implemented (i.e., $D_c=f(D)$ for some odd function $f$ which is the identity at low energy but tends to $\pm c$ for energies above $c$)

Integrating over the fermion matter modes gives a determinant, which is evaluated as the exponential of the trace of a logarithm. This logarithm is modified at high energies, essentially so that modes with energy greater than a cut-off energy $c$ do not contribute.

The asymptotic expansion of the determinant is well-known \cite{V,BS}, giving 
$$S_\text{eff}=\int\frac{-c^2N}{32\pi^2}-2\Lambda_0  + \frac{c N} {192\pi^2} R + \text{etc.} \quad\dd V,$$
with `etc.'~being an infinite series of terms which are higher order in the curvature. The Einstein-Hilbert term $R$ appears in this effective action with the correct sign (the Misner-Thorne-Wheeler sign convention is used, in which the gravitational action is $\int-2\Lambda+\frac1{16\pi G}R\;\dd V$ ). The fermions induce a large positive cosmological constant. As it is not clear how the cosmological constant behaves under renormalisation, it is not clear if this is a problem. However the bare constant $\Lambda_0$ can in any case be chosen to be negative and cancel it if necessary.  

%To render this finite, one actually calculates the ratio of  determinants, the second one being the same formula but for some fixed reference metric.  However, the details of the low-energy physics are not a concern in this paper, so this.

In this way one obtains an induced action for the metric. The coupling constants are not those measured at low energies, but should run towards them as the energy scale decreases from the Planck scale downwards due to  renormalisation. In particular, one expects the higher-order terms to have negligable effect at macroscopic scales.

\section{Induced standard model}
%\subsection{The Connes-Chamseddine spectral action}
Realistic matter fields for the standard model plus right-handed neutrinos coupled to gravity can be introduced by generalising this model problem. The fermion field $\Psi$ is now a vector whose components have indices both for spin and for the three generations of sixteen different particle types (including a right-handed neutrino)  The Dirac operator becomes schematically
$$D=D_G+A+yH$$
with $A$ the $U(1)\times\SU(2)\times \SU(3)$ gauge fields, $H$ the Higgs and $y$ the Yukawa coupling matrix.

Since the fermion functional integral will induce an effective bosonic action, the simplest possible, and most radical, hypothesis is that the bosonic term in the bare action is absent.
Thus this paper suggests a new bare action in which there is only a fermionic term, plus possibly a bare cosmological constant. 
%The formula is written here for the more natural Lorentzian version of the theory \cite{BSM}.
\begin{equation}\label{SMaction} I_{SM}=\int\left(\overline\Psi D\Psi - 2\Lambda_0\right)\dd V.\end{equation}

To use the existing literature, the problem of calculating the induced bosonic action $S_b$ will be considered in Euclidean signature. This should be directly related to the Lorentzian functional integral at least in the case of static space-times. Assuming the phase can be ignored, then the fermion functional integral yields
$$\det D=e^{-S_b}=e^{\frac12\tr\log D^2}.$$

A standard way to implement the cutoff at energy $c$ is to use the  exponential integral
$$\Ei(x)=\int_x^\infty\frac{\dd t}te^{-t}.$$
This function is damped exponentially for $x>>1$ but is asymptotic to $-\log x$ for $x<<1$

Thus $\log D^2$ is replaced by $-\Ei(D^2/c^2)$ and so the induced bosonic action is defined by
$$S_b=\frac12 \tr \;\Ei(D^2/c^2)$$
for this choice of cutoff. More precisely, to avoid infinite volume and other infra-red divergences one defines the difference with a standard reference geometry $D_0$ 
\begin{equation} \label{bosonicaction} S_b=\frac12 \tr \;\Ei(D^2/c^2)- \frac12 \tr \;\Ei(D_0^2/c^2)\end{equation}
but this additional constant will usually be ignored in the following for simplicity.

Of course we do not know if this form of the cut-off is realised in the true high-energy theory; it is just a convenient choice which illustrates the phenomenon of the induced bosonic action. In fact several features of the induced action are fairly insensitive to the choice of cut-off.

The fermionic part of \eqref{SMaction} admits an elegant description in terms of Connes' non-commutative geometry, explained in \cite{BSM} in Lorentzian signature and \cite{CSM,CCM} in Euclidean signature. However one does not need to understand this geometry for the following calculations.

The bosonic action for a very closely-related problem is also given in \cite{CC, CCM}. The  Connes-Chamseddine spectral action for a Euclidean $D$ is
$$ S_{CC}=\tr\;\chi(D^2/c^2)$$
with $\chi(x)$ a positive function that falls to zero at large $x$ such that the moments
$$\int_0^\infty\chi(u)u\,\dd u\quad\quad\text{and}\quad\quad\int_0^\infty\chi(u)\,\dd u$$
are finite. The difference to \eqref{bosonicaction} is that $\chi(0)$ is finite whereas $\Ei$ diverges. This means that our induced bosonic action $S_b$ will differ from the Connes-Chamseddine one at low energy, and this turns out to be important, as will be explained shortly.

The Connes-Chamseddine asymptotics is taking $c\to\infty$. Since $D$ contains fermion masses, which one wants to keep fixed, this should be thought of as scaling the cutoff function $\chi$. The resulting asymptotic formula for $ S_{CC}$ gives the complete bosonic part of the standard model coupled to gravity, including the Einstein-Hilbert action, the Yang-Mills action and the Higgs action with all the correct couplings and signs, and 
with some particular relations between the coupling constants. Most particularly, it gives the relation between gauge coupling constants
\begin{equation}\label{grand}\alpha_3= \alpha_2=\frac53\alpha_1\end{equation}
well-known from grand unification theories. There are also three other reasonably realistic relations between coupling constants, but these won't be analysed here.

%identifying the bosonic action at the highest energy scale with the renormalised action

Thus to analyse the asymptotics of our bosonic action $S_b$, it is sufficient to analyse the difference to the Connes-Chamseddine formula. It should be reasonably clear that again the standard model Lagrangian is obtained but with different coupling constants. In the following, the gauge coupling constants will be analysed, leaving the details of the Higgs sector for future work.

The induced Yang-Mills term for $S_b$ is well-known. For one Dirac fermion of mass $m$ the leading term is
\begin{equation}\label{YM}\frac 1{g^2}\int\frac12\,\tr\,F^2\;\dd V\end{equation}
with
$$\frac 1\alpha=\frac{4\pi}{g^2}=\frac 2{3\pi} T(R)\log c/m$$
and $T(R)=\frac12$ for the fundamental of $\SU(2)$ or $\SU(3)$ and the hypercharge squared for $\U(1)$.

Therefore the Yang-Mills coupling constant induced by a number of Weyl fermions (which contribute as one half of a Dirac fermion) is
\begin{equation}\label{YMSM}\frac 1\alpha=\frac 1{3\pi} \sum_{\Atop{\text{Weyl}}{\text{fermions}}}T(R)\log c/m\end{equation}

The Connes-Chamseddine formula for a Dirac fermion is, by contrast,
 \begin{equation}\label{YMCC}\frac 1{\alpha_{CC}}=\frac 1{3\pi} T(R)\chi(0).\end{equation} 

These formulas can be reconciled by the following observation. Equation \eqref{bosonicaction} would fall into the Connes-Chamseddine framework if the $\Ei$ function was modified to include an infra-red cutoff at an energy $r>m$, i.e. defining $\chi(x^2/c^2)=\Ei(x^2/c^2)$ for $x\ge r$ and taking the constant value below, $\chi(x^2/c^2)=\Ei(r^2/c^2)\simeq\log(c^2/r^2)$. Then the induced Yang-Mills term would be \eqref{YM} with
$$\frac 1\alpha= \frac 2{3\pi} T(R)\log c/r$$
which agrees with \eqref{YMCC}.

In fact it is clear why this difference occurs. In the Connes-Chamseddine limit $c\to\infty$, the infra-red scale $r$ also increases to infinity and will thus eventually exceed any fermion mass. %Therefore the Connes-Chamseddine asymptotic spectral action only agrees with the induced action defined here in the high-energy part. 

\section{Unification}

The question to be settled is whether the bold Ansatz \eqref{SMaction} is compatible with known particle physics. Since this is the bare action, it would not be manifest in that form at low energy, i.e., significantly below the cut-off. So the question is, can one see it by a suitable running to high energies?

The conventional unification picture is use the running of the renormalised gauge coupling constants $\widehat\alpha$ to compute the gauge coupling constants at some high energy $\mu$ starting from measured values at energy $\mu_0$. At one loop,
\begin{equation}\label{SMrunning}\frac1{\widehat\alpha(\mu)}= \frac1{\widehat\alpha(\mu_0)}+\frac 1{6\pi}\left(11 T(A)- \sum_{\Atop{\text{Weyl}}{\text{fermions}}}2T(R)- 
 \sum_{\Atop{\text{Higgs}}{\text{scalars}}}T(R)\right)\log\mu/\mu_0\end{equation}
with $T(A)$ the group theory factor for the adjoint of the gauge group, equal to $3$ for $\SU(2)$, $2$ for $\SU(2)$ and $0$ for $\U(1)$. Starting with measured data at $\mu_0=M_Z$, the running is as shown in figure 1. The couplings fail to meet at the values \eqref{grand} predicted by grand unification and also by the Connes-Chamseddine spectral action,
identifying the bosonic action at the highest energy scale with the renormalised action. This is usually taken to suggest that there must be some new physics.

\begin{figure}$$\epsfbox{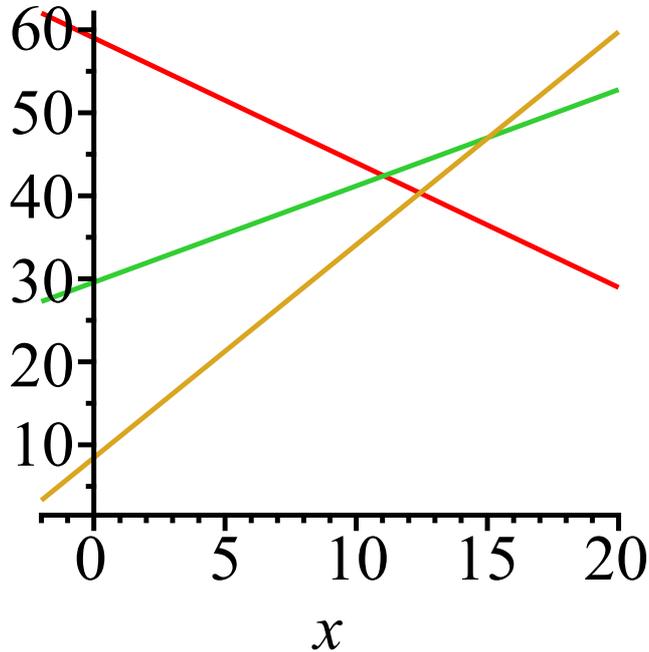}$$
\caption{Standard model running of gauge couplings. Horizontal axis $x=\log_{10}\mu/\mu_0$. Vertical axis $3/(5\widehat\alpha_1$), $1/\widehat\alpha_2$, $1/\widehat\alpha_3$.}
\end{figure}

This type of running could not possibly account for \eqref{SMaction}. The vanishing of the bare bosonic action means that it has infinite bare values of both the gauge and gravitational coupling constants, and is thus inaccessible to perturbation theory. Clearly it makes no sense to ignore the vanishing of the gravitational part of the action and just run the gauge coupling constants, as in \eqref{SMrunning}.

A different test of \eqref{SMaction} is suggested by the induced bosonic action. This is obtained by integrating out some of the fermion degrees of freedom, and can be carried out to give an effective field theory which can be used below the mass of the fermions that are integrated out.

 Thus it suggests looking at the effective theory obtained in this way, with gauge coupling constants $\alpha_b$. These are
 computed using the theory of fermion decoupling\cite{MM}. According to this theory, in a simplified model with just one fermion, the coupling constants match at the fermion mass
$$\widehat\alpha_b(m)=\widehat\alpha(m),$$
with $\widehat\alpha$ being the coupling for the full theory.
Due to the different running, they do not match at other scales.
 
Starting at low-energy scales, it is possible to run the gauge coupling constants to high energy because one should meet not the zero bosonic action as in the previous case, but now the induced bosonic action. In this theory the gauge and gravitational couplings are not zero. In fact, it makes sense to use perturbation theory, and since the Einstein-Hilbert action does not vanish, it is not completely unreasonable to ignore any interaction with gravity in the running of the gauge couplings.

The first point is that although this requires heavy fermions this does not require exotic new physics. In fact the seesaw mechanism used to generate neutrino masses already has  very heavy neutrinos which would be expected to induce some sort of bosonic action. Further, one could easily postulate an entire additional standard-model generation of as yet unobserved massive fermions.

Taking this scenario leads to an induced bosonic action that is very similar to the Connes-Chamseddine spectral action. The coupling constants are given by \eqref{YMSM}, which is not necessarily the grand unification relation \eqref{grand}.

In fact with an entire generation of massive fermions with equal masses $M$, one obtains the induced bosonic action
\begin{equation}\label{equalM}\frac 1\alpha=\frac 1{3\pi}\log (c/M)\; \sum_{\Atop{\text{Weyl}}{\text{fermions}}}T(R),\end{equation}
specialised from \eqref{YMSM}. The sum over group theory factors  leads directly to the grand unification relation \eqref{grand}. This special case therefore leads to the Connes-Chamseddine bosonic action.

\section{State sum models}

A promising approach to a fundamental theory is to use the mathematics of category theory, which in recent years has shown a remarkable relation with topological quantum field theory, and has also been used to construct models of quantum gravity without matter.
Although the state sum models are discrete in nature, it is envisaged that an approximate continuum description should emerge at energies below the Planck scale. Therefore state sum models are constructed with this limit in mind - it guides the expectations of the physical content of the model.

A state sum model on a $d$-dimensional manifold is constructed given a certain type of weak $d-1$--category. This category has a set of objects, or 0-morphisms. Then there is a set of 1-morphisms which map between the 0-morphisms, 2-morphisms which map between pairs of 1-morphisms that map between the same 0-morphisms, 3-morphisms between 2-morphisms, ... and so on, up to a set of $d-1$-morphisms. These morphisms have $d-1$ different composition laws (one for each level) and the axioms for these, and relations between them, increase in complexity quite dramatically as $d$ increases.

The partition function of the state sum models is defined using some convenient combinatorial decomposition of the $d$-manifold, most commonly using a triangulation. A \emph{state} is an assignment (or labelling) of a $k$-morphism from the category to each $k$-simplex of the manifold, for all $k$ from 0 up to $d-1$. These have to satisfy an obvious consistency condition. For example, the 2-morphism labelling a triangle has to be a mapping from the 1-morphism labelling one of the edges to the 1-morphism labelling the product of the other two edges.

These are the physical variables, and the discrete version of the functional integal, called the partition function, is computed by a sum over a suitable basis of these variables. The summand is a suitable \emph{weight} factor (a complex number) defined as a product of terms which are each local in the variables (i.e., depend only on the variables in one simplex). Examples of state sum models from physics include
\begin{itemize}
\item  lattice gauge theory
\item 2d Yang-Mills
\item 3d quantum gravity
\item 4d quantum gravity models 
\end{itemize}

These examples do not use all of the complexity of $n$-categories. For example, a group determines an $n$-category for all $n\ge1$ with one object and its 1-morphisms the group elements. The higher morphisms are all just the identity mapping. In this way, one can have a lattice gauge theory in any dimension $d\ge2$. 

The fundamental example in four dimensions is the Crane-Yetter model\cite{CKY} constructed using the braided monoidal category of representations of a quantised Lie algebra. This is a tricategory (a weak 3-category) in which there is only one object and one 1-morphism. Thus in this example, a state is specified by labelling triangles and tetrahedra with data from the 
braided monoidal category. One expects that there should be many other interesting examples of tricategories which lead to interesting state sum models in four dimensions in which there are non-trivial 1-morphisms. However the stock of known suitable monoidal bicategories is as yet somewhat limited. 

The Crane-Yetter model is a topological gauge theory, obtained by deforming the Ooguri model\cite{O} which integrates over all flat connections. Models of quantum gravity have been constructed by modifying Crane-Yetter models in various ways. The experience with this, and similar models on 3-manifolds, leads to the following general list of properties of the state sum models which may realise physical models. These features have all been seen in various models and it is not unreasonable to expect there to exist state sum models with all of them at once. The wish-list of properties for a state sum model is

\begin{itemize}
\item It defines a diffeomorphism-invariant quantum field theory on each 4-manifold
\item The state sum can be interpreted as a sum over geometries
\item Each geometry is discrete on the Planck scale
\item The coupling to matter fields can be defined
\item Matter modes are cut off at the Planck scale
\item The action can include a cosmological constant
\end{itemize}

Diffeomorphism invariance here actually means invariance under piecewise-linear homeomorphisms, but this is essentially equivalent. The piecewise-linear homeomorphisms are maps which are linear if the triangulations are subdivided sufficiently and play the same role as diffeomorphisms in a theory with smooth manifolds. This invariance is seen in the Crane-Yetter model and also in the 3d gravity models, the Ponzano-Regge model and the Turaev-Viro model, the latter having a cosmological constant. The 3d gravity models can be interpreted as a sum over geometries, a feature which is carried over to the four-dimensional gravity models \cite{BC,EPRL,FK}, which however do not respect diffeomorphism invariance.

The coupling of the 3d gravity models to matter is studied in \cite{BO,FL}, and extended to 4d models in \cite{BF}. A model with a fermionic functional integral have been studied in \cite{FB,FD}, though as yet there is no model which respects diffeomorphism invariance. This is clearly an important area for future study.

The most obvious omission from this list is the ability to implement the Einstein-Hilbert action. In fact, experience with state sum models in four dimensions so far is that there are models with diffeomorphism-invariance but no Einstein-Hilbert action, and there are models implementing the Einstein-Hilbert action but having (at best) only approximate diffeomorphism-invariance. Thererfore it is obviously important to question whether this is a fundamental problem, or merely the result of not including matter in quantum gravity  models. If the gravity action is induced by the matter degrees of freedom, then the problem can be completely circumvented, or rather turned into the problem of constructing state sum models with both gravitational variables and realistic matter degrees of freedom.

\section{Conclusion}
This paper investigates that idea that the bosonic action of the standard model coupled to gravity should be induced from the bare fermionic action, and places this in the context of state sum models where an induced action should perhaps naturally arise. 

A bosonic action similar to the Connes-Chamseddine spectral action arises from integrating out massive fermions. One of the merits of this proposal is that it gives an explanation for the spectral nature of the bosonic action; it is due to the fact that the fermion determinant is spectral. It also allows modifications to the Connes-Chamseddine spectral action when the massive fermions do not all have the same mass.

\section*{Acknowledgement} Thanks are due to the hospitality of the Alain Connes at the IHES, where these ideas were first presented in December 2009 at a meeting supported by the QG network of the European Science Foundation. Thanks are also due to discussions with Christoph Stephan and Harold Steinacker.

\end{document}